\documentclass{article}

\usepackage[a4paper, total={6in, 10in}]{geometry}
\usepackage[T1]{fontenc}
\usepackage[hidelinks]{hyperref}
\usepackage{amsmath,amssymb,amsfonts}
\usepackage{algorithmic}
\usepackage{graphicx}
\usepackage{textcomp}
\usepackage{xcolor}
\usepackage{graphicx}
\usepackage[english]{babel}
\usepackage[utf8]{inputenc}
\usepackage{longtable}
\usepackage{fancyhdr}
\usepackage{color}
\usepackage{url}
\usepackage{enumitem}
\usepackage{multirow}
\usepackage{setspace}
\usepackage{caption}
\usepackage{subcaption}
\usepackage{url}
\usepackage[font=bf,font=small]{caption}
\usepackage{wrapfig}
\usepackage{makecell}
\usepackage{array}
\makeatletter
\newcolumntype{"}{@{\hskip\tabcolsep\vrule width 2pt\hskip\tabcolsep}}
\makeatother

\def\BibTeX{{\rm B\kern-.05em{\sc i\kern-.025em b}\kern-.08em
    T\kern-.1667em\lower.7ex\hbox{E}\kern-.125emX}}

\graphicspath{{img/}}

\newcommand{\myparagraph}[1]{\textbf{#1}.~}
\newcommand{\new}[1]{#1}
\newcommand{\iosif}[1]{#1}
\newcommand{\myemail}[1]{\href{mailto:#1}{#1}}

\begin{document}
\title{Anomaly Detection Within\\
Mission-Critical Call Processing
}
%
%
\author{\hfill Sean Doris$^1$\hfill Iosif Salem$^2$ \hfill Stefan Schmid$^2\thanks{Supported by German Research Foundation (DFG) project ReNO (SPP 2378), 2023-2027.}$\hfill\\~\\
%
%
$^1$Motorola Solutions, Inc., 2600 Glostrup, Denmark\\
\myemail{sean.doris@motorolasolutions.com}
\and 
$^2$TU Berlin, 10587 Berlin , Germany\\
\myemail{iosif.salem@inet.tu-berlin.de}, \myemail{stefan.schmid@tu-berlin.de}}
\date{}
%
%
%
\maketitle              
\begin{abstract}
With increasingly larger and more complex telecommunication networks, there is a need for improved monitoring and reliability. 
Requirements increase further when working with mission-critical systems requiring stable operations to meet precise design and client requirements while maintaining high availability. 

This paper proposes a novel methodology for developing a machine learning model that can assist in maintaining availability (through anomaly detection) for client-server communications in mission-critical systems. To that end, we validate our methodology for training models based on data classified according to client performance. The proposed methodology evaluates the use of machine learning to perform anomaly detection of a single virtualized server loaded with simulated network traffic (using SIPp) with media calls.
The collected data for the models are classified based on the round trip time performance experienced on the client side to determine if the trained models can detect anomalous client side performance only using key performance indicators available on the server. 

We compared the performance of seven different machine learning models by testing different trained and untrained test stressor scenarios. In the comparison, five models achieved an F1-score above 0.99 for the trained test scenarios. Random Forest was the only model able to attain an F1-score above 0.9 for all untrained test scenarios with the lowest being 0.980. The results suggest that it is possible to generate accurate anomaly detection to evaluate degraded client-side performance. 

\end{abstract}
\section{Introduction}
\label{sec:intro}
Anomaly detection has become a key focus for many industries as a method of producing accurate monitoring and analysis. Their applications range from assisting medical diagnosing and establishing a prognosis of cancer 
\cite{wong_machine_2018} 
to monitoring the system health of servers and routers in various telecommunication applications 
\cite{jin_accurate_2016}.
While the applications differ, many anomaly detection applications are interested in being used in mission-critical applications where safety and reliability are essential. The use of machine learning (ML) in these scenarios allows for improved data analysis by detecting complex patterns that could have otherwise gone unnoticed and could be used as predictors for future failures.

Mission-critical telecommunication systems have become increasingly advanced to handle the growing demand for higher data throughput and scalability.
New features and bug fixes are regularly being deployed through releases of updates putting systems in a state of flux.
Scalability is often implemented using several multi-card chassis, each having numerous virtual machines or containers to handle call volume \cite{romanov_evaluation_2020}. All the while, the systems must be thoroughly tested to ensure that telecommunication systems have a high degree of reliability to ensure regular availability. Various potential failures can occur on these systems, such as reduced link quality, resource contention, orphan processes, hardware failures, and thermal limiting. Even minor hard-to-detect failures could cause performance degradation beyond safe operating conditions. Furthermore, when one virtualized instance fails, it can affect the performance of other connected virtualized instances by increasing response times and delays \cite{cisco_networking_academy_program_connecting_2014}. This degradation can be difficult to diagnose in a large, complex, modularized network. Modularizing systems using virtual machines and docker containers allows for improved maintainability and load balancing
\cite{azizi_energy-efficient_2020}. 

Mission-critical applications' uptime requirements can require a 99.999\% uptime with no more than 6 minutes of downtime per year\cite{avdotin_ofdma_2019}. These mission-critical systems are often used in emergency response to support emergency workers to safely and quickly reach and provide the required assistance to those in need. As such, mission-critical telecommunications require effective methods to help ensure the system's stability can meet the performance required by its applications. 
For real-time applications, downtime also includes periods when performance is degraded beyond the levels of its service requirements. Due to the limited downtime or anomalous instances, it would be more efficient to develop a method generating simulated anomalous data than trying to collect anomalous instances from active systems.
Additionally, by providing autonomous monitoring for each virtualized system, it can reduce the complexity of applying self-healing methods since fixes can be applied without affecting the larger system 
\cite{liu_case_2008}.

The use of ML to detect faults such as intrusion, DDOS attacks, memory or hardware faults is an area of focus in many research efforts today 
\cite{khalil_machine_2020,tuncer_online_2019}.
Another application of ML models is that they can assist in maintaining the uptime in mission-critical systems by training a model that can detect if a system's performance has degraded beyond acceptable specifications. These specifications change from application to application; however, within call processing, this can be the server's processing time required to set up the communication. This paper aims to create a ML model that accurately detects abnormal performance degradation that exceeds the normal operating parameters of a virtualized system. 

\myparagraph{Scope}
This paper describes a procedure for simulating call processing under different scenarios with varying loads. 
The call processing setup is two virtual machines loaded with SIPp call traffic with media (SIPp simulates the SIP call and messaging protocol \cite{gayraud_sipp_2014}). One virtual machine will be the server, and the other will be the client. The two virtual machines will exist on the same host system, with all communications taking place over the local building network. A definition of degraded performance and how it is classified will be determined by analyzing the normal operating behaviour of the test setup. By defining the specifications of what is expected for the system, a level can be set for what resulting performance on the client side is classified as anomalous or non-anomalous.

Various stressors will be applied to the system using stress-ng (open source software for stressing systems \cite{king_stress-ng_2022}). Each type of stressor shall have two different load levels applied, a low level that will not result in degraded performance on the client side and one that will. 
The data is classified as anomalous if the round trip time (RTT) reaches anomalous levels due to performance degradation caused by stressors active on the server side. The RTT consists of the client-server-client communication delay and the server processing time and we will assume reliable client-server communication. Thus, increased RTT indicates increased server processing time due to anomalous server behaviour. While other client or design parameters could be used for classification, RTT is selected as a classifier for validation as it is a metric that could be applied to many mission-critical approaches in communication systems. In theory, the methodology of using measurable design parameters for classifying training data could be adapted to other mission-critical systems by using a different set of more relevant parameters. 

The ML models are tested and trained with data from measuring key performance indicators (KPI) on the server side during different scenarios, we will build a training and test data set for the ML models. The scenarios being tested are (i) unstressed scenarios when no stressor is active, (ii) non-anomalous stressed scenarios when a stressor is active, but the RTT does not exceed anomalous levels and (iii) anomalous scenarios where the stressors will cause the RTT to exceed anomalous levels. The models will be tested with untrained scenarios where the data from the stressors are exclusively used for testing. The untrained scenarios will validate that the ML models are accurate for other anomalous scenarios and are not overtrained for the stressors that they were trained with. 

From the results, seven well-known ML models were compared over various scenarios using the stressors tested. The diverse types of ML models that were selected include support vector machines, tree-based, cluster-based, and probability-based models. In addition, both classification and regression, as well as both supervised and unsupervised methods, are tested to determine if there are any distinctions in their ability to be used for anomaly detection under this methodology and the generality of the approach.

\myparagraph{Contributions}
Our main contributions are the following:

\noindent 1)~~We leverage client-server RTT in the call-setup protocol as the main classifier for defining anomalies and (non-)anomalous scenarios.
The novelty of this approach is that it more accurately classifies anomalies compared to the standard classification of data collected from when the system is stressed or unstressed, since low-stress scenarios may still be tolerated by the call-processing system.


\noindent 2)~~We generate accurate anomaly detection that can evaluate degraded client-side performance.
All tested supervised ML models achieved F1-scores above 0.99 for trained test scenarios excluding Gaussian Naive Bayes. However, many of the supervised models had reduced performance when comparing the machine-learning results with the stressor test results. Random Forest performed the best across all untrained test scenarios, with the lowest result being 0.980.

\myparagraph{Outline}
Section \ref{sec:background} will provide an overview of 
the measurement and anomaly generation applications. 
Section \ref{sec:methodology} will provide the methodology for the paper. Section~\ref{sec:results} will display and discuss the results of the performance for ML models on the tested scenarios. Finally,  in section \ref{sec:relatedwork}, we present related work. 

\section{Background}
\label{sec:background}
\label{sec:anomalygen}
Training ML models require applications to measure and store the KPIs being measured on the system. Generating the data for training the model also requires a system that is under load to monitor and another application to create an anomalous state to generate data for training the supervised ML models. 

\myparagraph{Measurement Tools \& Applications}
This section will denote what tools will be used to measure the KPI required for anomaly detection.
%
The selected tools are \texttt{vmstat}, \texttt{iostat}, and \texttt{netstat}. 
\texttt{vmstat} is an application  can be used to measure various hardware performance metrics on the virtual machine\cite{ware_vmstat8_nodate}. 
It provides access to active/inactive memory, context switching, CPU interrupts, and buffer memory usage.
\texttt{iostat} command can be used on Linux systems to monitor the IO usage of connected devices\cite{godard_iostat1_nodate}. 
\texttt{netstat} is a command that can be used to display network related metrics for Linux based systems. \texttt{netstat} can display a large amount of detailed information regarding the system\cite{welsh_netstat8_nodate}. 


\myparagraph{Anomaly Generation Application}
We used \texttt{stress-ng} to generate anomalous data on the systems to be tested. 
%
\texttt{stress-ng} is open-source software that can stress a system in numerous ways through its selection of over 280 selectable and customizable stress tests\cite{king_stress-ng_2022}. The tests can be selected to stress targeted pieces of a system. This can be ideal for generating anomalous data for select KPIs to ensure all of the features we are measuring can have anomalous training data that can stand out from a healthy system measurement. Several authors have researched ML using anomalous data generated using \texttt{stress-ng} to simulate various instances such as CPU, memory, and disk faults\cite{gulenko_detecting_2018,gulenko_evaluating_2016,casas_machine-learning_2016,sauvanaud_towards_2016,samir_detecting_2019}. 

\myparagraph{SIPp}
SIPp is a software that can simulate and allow performance testing of SIP protocol\cite{gayraud_sipp_2014}. Using SIPp on two different platforms, one can have user agent client (UAC) and user agent server (UAS) communication using different packet types. SIPp can perform operations during a call, such as executing commands or streaming multimedia such as audio. During communication, it can provide detailed statistics such as RTT, call rate, and other call-related statistics. These statistics allow a user to determine the quality of a given communication. 
%
We will use SIPp as the method of producing system and network load on our virtual machine and measuring the RTT for all communication during a period.

\section{\iosif{Data Collection Methodology}}
\label{sec:methodology}

This section of the paper will outline the steps and methods used for the data collection. The data collection will include the normal and anomalous system data for training the ML models used in this project. The platform used is a VM running using two cores from an Intel Xeon E5-2670, 3GB of memory, and a virtual operating system running Red Hat Enterprise Linux 8.4 (Figure \ref{fig:test_setup}).

\subsection{SIPp}
The call scenario that was used is a customized version of the default user agent client packet capture (UAC PCAP) call scenario offered by SIPp. The PCAP scenario simulates media communication between the client and server end. This media communication generates significantly more load than the standard UAC scenario, which has a small exchange of packets. Additionally, the load can be a better simulation of an actual scenario, as in the UAC PCAP scenario, data is streamed between the two. For the call scenario, a call rate of 70 call/s provided a sufficiently high load on the server side but not high enough to cause the server to be overloaded and lead to timeouts due to excess load. 

\begin{figure*}
\centering
\begin{minipage}{.3\textwidth}
  \centering
  \includegraphics[width=.68\linewidth]{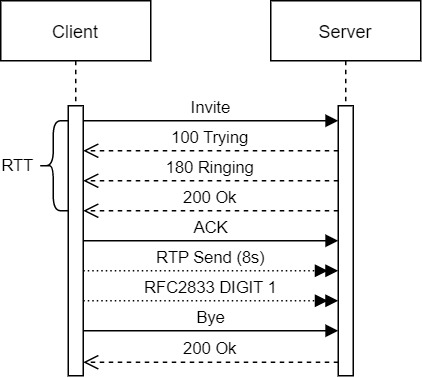}
  \captionof{figure}{UAC PCAP call\newline scenario}
  \label{fig:call_scenario}
\end{minipage}%
\begin{minipage}{.3\textwidth}
  \centering
    \includegraphics[width=0.75\linewidth]{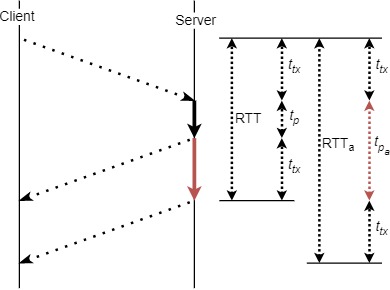}
    \captionof{figure}{Round trip time\newline (RTT) diagram}
    \label{fig:RTT}
\end{minipage}%
\begin{minipage}{.3\textwidth}
  \centering
  \includegraphics[width=0.65\linewidth]{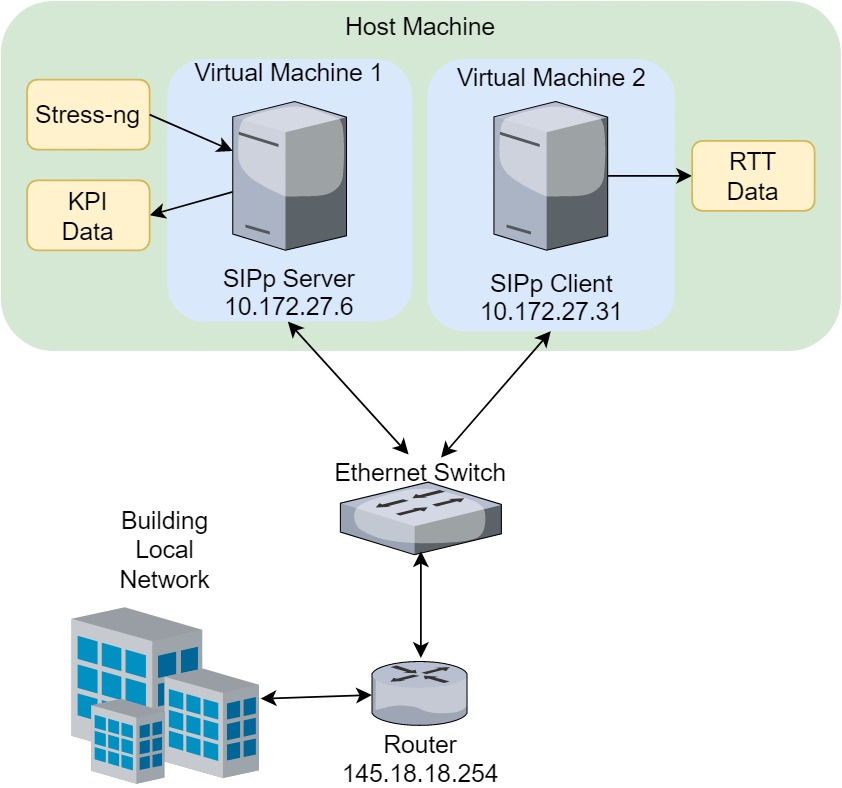}
  \captionof{figure}{Diagram of the\newline test setup}
    \label{fig:test_setup}
\end{minipage}%
\end{figure*}

Settings that were adjusted within the default UAC PCAP scenario were the timeout settings that needed to be adjusted to prevent calls from staying open and limiting the startup of new calls. Otherwise, the call rate for the scenario could be unstable. The retransmission time for the invite message was reduced from 500ms to 50ms due to the network's very short latency. The default 500ms caused large spikes in the average due to the long retransmission delay; reducing the retransmission time to 50ms reduces these spikes but still makes them impactful. The RTT is measured from the initiator's or client's end, starting the RTT timer on the invite message and stopping it upon receiving all set-up messages just before the initiator sends the ACK (figure \ref{fig:call_scenario}).
\newcommand{\jointfigsize}{}


RTT can measure the quality of communication traffic, especially in mission-critical systems where if RTT spikes are left unaccounted for, they could lead to disruptions. Under the assumption that the network connecting two systems is stable and transmission time ($t_{tx}$) remains unchanged, changes in RTT can be used to quantify the changes in the time to process ($t_p$) a message provided to the server before replying (Figure \ref{fig:RTT}). Suppose the connection of a system is stable and the RTT is increased. In that case, the increased RTT is due to increased processing time. This increase in processing time can be due to the server being in an anomalous state ($t_{p_a}$), increasing the server's time to respond. Monitoring the server's KPIs while measuring the RTT, one can classify the RTT from the client's end as anomalous or non-anomalous and apply that same classification to the KPIs logged on the server end. With this application, one can build enough data to train and test ML models to monitor the KPIs of the server. This claim of a stable network while monitoring a system under a load of SIPp voice calls is similar to other studies that used a similar methodology for assessing the viability of containers \cite{romanov_evaluation_2020} or studying anomaly detection \cite{sauvanaud_towards_2016}. \new{Under stable network conditions data can be collected to train ML models that could be applied to similar active systems where these assumptions aren't present and RTT cannot be used alone as a suitable metric for determining system stability.}


\subsection{Anomalous State}

A definition for what is classified as anomalous must be made before determining what is anomalous and the acceptable RTT. An anomalous state should have an RTT that exceeds the levels expected for transmission. Unexpectedly high RTT could cause degraded communication quality experienced by the client. With such a definition, ML models can monitor and ensure a system's reliability. 

To determine what value of RTT is anomalous, the RTT of an unstressed system running the UAC PCAP scenario is measured for 20 hours. The RTT is measured from the client's end to evaluate the quality of the communication from the client's perspective. Meanwhile, the KPIs are measured on the server end. The RTT is averaged over the time frame of the data point in the KPI log so that the averaged RTT can be used to classify the data point. Mathematically, the classification for what is outside normal distribution is any RTT that exceeds the third standard deviation for the averaged RTT data points over the full 20-hour measurement time period.

To measure the effectiveness of the ML models, the metrics that are used are accuracy 
(eqn. \ref{eq1}) 
and F1 score. 
(eqn. \ref{eq4}). 
To perform these calculations, the classification matrices need to be defined for the provided test scenarios that are performed. For any tests where the expected RTT level is below the anomalous RTT level, a positive result is the model correctly predicting a non-anomalous result. For any tests where the expected RTT is above or equal to the anomalous RTT level, a positive result is the model correctly predicting the system's state as anomalous. 
The detailed classification matrices are in the appendix \ref{Anomaly-classification}.

\subsection{Stress test selection}
The following tests were selected for creating data that will stress the measured KPI of the system and create both anomalous and non-anomalous data for different segments of the system. The collected data from these tests are used for training and testing of the models:

\begin{itemize}
    \item \textbf{CPU} to cause CPU load, stress-ng iterates through a list of loads such as floating point calculations, square root and other computational loads such as the eight queens problem used in computational load tests
    \item \textbf{icache} to simulate a load causing interference in the instruction cache leading to further instruction stalls and delays to the pipeline
    \item \textbf{aio} to cause stress on the io by triggering several small writes and reads to the disk
    \item \textbf{UDP} to cause stress on the local network sockets for UDP
    \item \textbf{rawsock} to create additional stress on the TCP/IP using the raw sockets on the local host
\end{itemize}

In addition to these stress methods, the four additional stressors are selected. These stressors will affect similar KPIs but will strictly be used to test the ML models and not to train them. With this data, the anomaly detection accuracy can be tested for other similar stressing scenarios. These will test for overtraining and ensure that models can classify any similar stimuli that may cause the system to enter an anomalous state. The chosen untrained stressors are:

\begin{itemize}
    \item \textbf{matrix} this iterates through several matrix operations. By enabling the yx option, the stressor can cause both CPU and cache interference
    \item \textbf{revio} has workers writing to temporary files on the hard disk. The writes are performed in reverse order. This is similar stress usage to aio; however, revio only performs writes 
    \item \textbf{rawudp} similar to rawsock, but instead it will stress using UDP 
    \item \textbf{rawpkt} sends and receives packets over the localhosts ethernet port
\end{itemize}

The provided tests are completed using different parameters to allow them to create enough stress to achieve anomalous RTT, as well as create a low-stress version that will have a non-anomalous RTT when performing non-anomalous data generation. The universal parameters for each stressor are the number of workers, and runtime and deadline parameters can be adjusted to cause different load levels.
There are a few other stressor-specific parameters that are changed, such as enabling the yx option in the matrix stressor to cause cache interference.

The resulting data from testing shall have at least 97\% of the generated data points as anomalous or non-anomalous depending on whether it is an anomalous test or non-anomalous test. The amount of variation in the average RTT for each test is different due to the different stressors. However, for non-anomalous stress loads, the RTT 
shall also be below the anomalous RTT threshold.

Other stress test options from stress-ng were tested. However, among the tests selected, they were also selected for being reproducible by producing RTT above the anomalous threshold consistently. 

\section{\iosif{Evaluation}}
\label{sec:results}

We present the results from data collection and methodology described in Section \ref{sec:methodology}. We present the test results from determining normal RTT with the applied system, the application of the ML models, analysis of the applied stress tests and how the ML models applied to the system react to the stressors.

\subsection{Determining Normal Conditions}

The calculation of the anomalous threshold was completed by collecting over 20 hours of system KPI and RTT data while SIPp was active. A call rate of 70/s was selected since it produced a relatively stable connection while generating enough load on the system. 
The 20 hours of test data were collected in two 10-hour intervals, with the systems being fully reset between the test intervals. 

The chosen frame time for the logging was 6 seconds, allowing enough time to execute the 2s monitoring intervals required for both \texttt{vmstat} and \texttt{iostat} as well as having minimal impact on the system KPI. The received RTT data was then parsed and averaged over the 6s logging period so each logging frame would have a correlated averaged RTT. A shorter logging interval could be selected if a mission-critical application required a quicker reaction time to an anomalous state, however, this would lead to a larger impact on system performance which could be an issue for systems with limited hardware.

From the measured results, the majority of the RTT ended up being less than 25ms, with averages over the frame being between 4-5ms.
Some of the spikes in RTT could be explained by network congestion, as the tests are not performed on a closed network. 
Due to the bursty nature of RTT there were several measurements exceeding 50ms, however, averaging the RTT for the duration of the frame significantly lowered the variance of RTT (Tbl. \ref{rtt_table}). The RTT around 50ms are caused by retransmissions where the sender didn't receive the ACK packet within the 50ms timeout and would then resend the packet.

\begin{table}[t!]\centering\scriptsize
\captionof{table}{RTT metrics from unstressed system runs}
\begin{tabular}{ |p{6cm}|p{1.2cm}|p{1.2cm}|p{1.2cm}|  }
 \hline
 \multicolumn{4}{|c|}{Unstressed RTT Metrics} \\
 \hline
 Metric & 1st run & 2nd run & Overall\\
 \hline
Average over full duration & 4.689ms & 4.662ms & 4.675ms\\
\hline
Standard deviation using rtt & 1.991ms & 1.892ms & 1.942ms\\
\hline
3rd standard deviation using rtt & 10.663ms & 10.339ms & 10.503ms \\
\hline
Standard deviation using rtt for the frame & 1.339ms & 1.372ms & 1.355ms\\
\hline
3rd standard deviation using rtt for the frame & 8.707ms & 8.776ms & 8.741ms \\
 \hline
\end{tabular}
\label{rtt_table}
\end{table}


By using the third standard deviation for the RTT of a frame from the client side and the KPIs that were taken during that same 6-second frame interval on the server side, a given KPI for that frame can be classified as normal or anomalous to provide training data for the ML models.

\subsection{Applying ML Models}


We describe the process of tuning of the parameters of the ML models we used. All collected data was pre-processed using linear scalar without using the mean to center the data before the scaling. The libSVM package offered through Python was used for the support vector machine methods\cite{CC01a}. The polynomial kernel was used for both Support Vector Classifier (SVC) \cite{aurelien_handsmachine_2017} and $\nu$-Support Vector Regression ($\nu$-SVR) \cite{scholkopf_new_2000}, while One Class Support Vector Machines (OC-SVM) \cite{tao_kernel_2020} performed best using the radial basis function (RBF) kernel. Due to the unequal number of non-anomalous cases in the training data, an additional non-anomalous test from the unstressed test case weight calculation was enabled to correct this offset.
OC-SVM was trained using only non-anomalous data and was tuned by adjusting the $\nu$ parameter to adjust the upper and lower bounds. A balanced approach was chosen to accurately detect anomalous instances while limiting the number of false positives occurring during non-anomalous data.

The Gaussian Naive Bayes (GNB) \cite{zhang_optimality_nodate}, k-Nearest Neighbors (kNN) \cite{barber_bayesian_2012},  Decision Tree (DT) \cite{segaran_programming_2007}, and Random Forest (RF) \cite{aurelien_handsmachine_2017} models were all applied using crates (compilation units) available in Rust programming language. GNB and DT are available through the Linfa-Trees and Linfa-Bayes crates(v.0.6.0), while kNN and RF are available in the Smartcore crate (v.0.2.1) \cite{noauthor_cratesio_2022}. 
The GNB model was tuned by adjusting the variance smoothing, the variance smoothing was increased from the default to add a portion of the feature with the largest variance to the stability equation.
When training the kNN model, the \emph{k} parameter for selecting the number of clusters was tested with \emph{k} values between 3 and 25. The number of clusters was tested using Euclidian, Hamming, Manhattan, and Minkowski distance metrics and found that Manhattan performed the best. The cover tree search algorithm was selected due to its increased speed and efficiency, it also improved accuracy compared to the brute force approach of Linear Search.

Tuning the tree-based models required adjusting the parameters for the split decision-making criteria. Entropy and Gini splitting criteria are available with these rust crates. Gini impurity measures the probability of a data point being misclassified if inserted into a randomly selected point. Entropy measures the disorder or variable composition of a provided node. A split is decided based on which split will decrease the entropy the most and thus increase the level of information the split provides. Both were tested on DT and RF, DT benefited more from the Entropy criterion while RF performance increased with Gini splitting. 
Both RF and DT had decision making parameters tuned to adjust the weight required to make a split. The default weights were low given the size of our dataset and would result in specialized nodes and trees with larger depths that proved impractical, leading to misclassifying some of the test data. 

\subsection{Model Results}

\begin{wrapfigure}{r}{0.47\textwidth}
    \vspace{-18pt}
    \centering
    \includegraphics[scale=0.12]{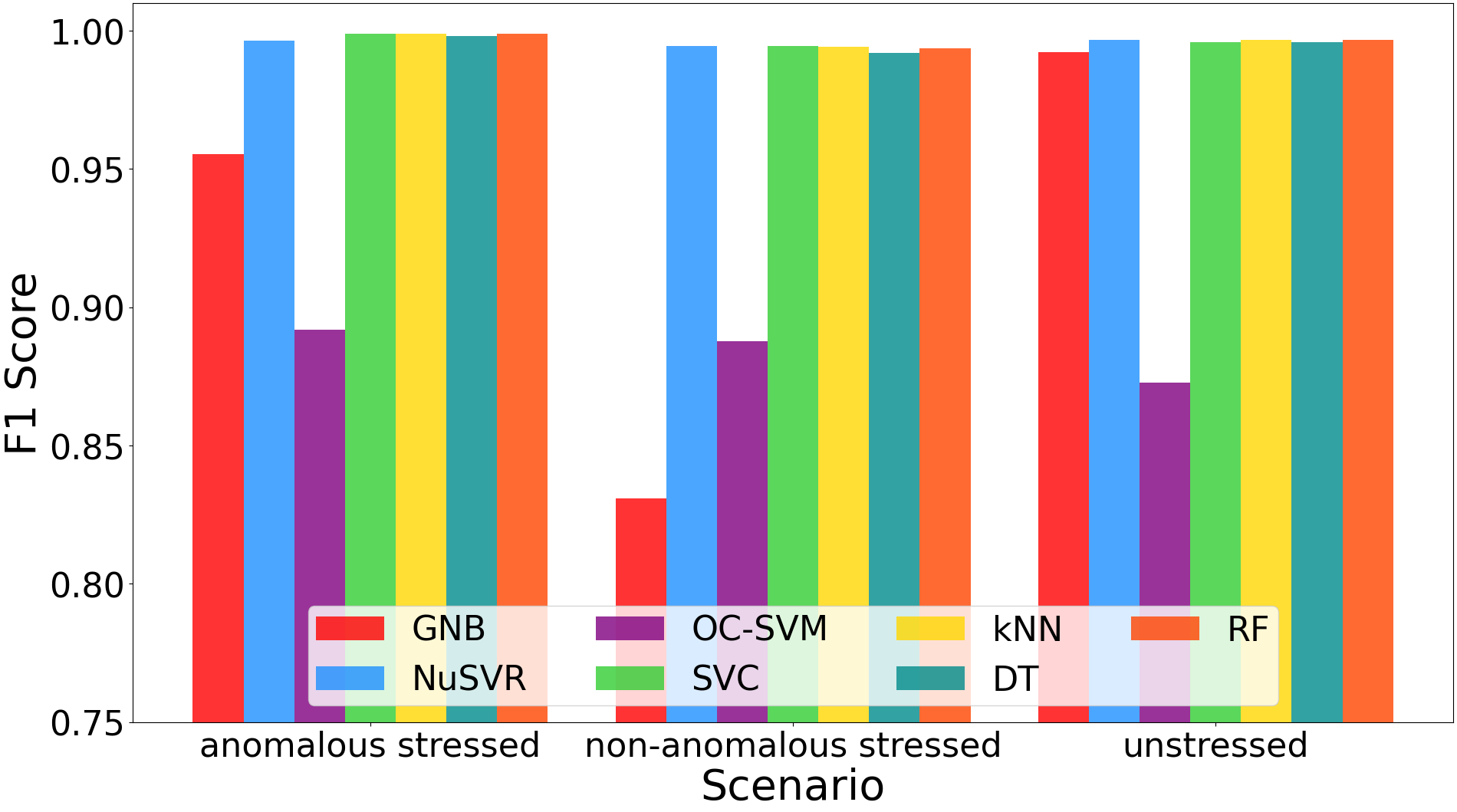}
    \caption{F1 score of models for different test scenarios}
    \label{M-F1-overall}
    \vspace{-10pt}
\end{wrapfigure}
Overall, all models produced an F1 score above 0.8 for unstressed, anomalous, and non-anomalous stressed scenarios. With $\nu$-SVR, SVC, kNN, DT, and RF producing F1-scores scores above 0.99. The RF model has the best F1-score for anomalous scenarios resulting in an overall F1 Score of 0.99898. $\nu$-SVR, kNN, and RF are tied for best overall F1 score for unstressed scenarios with a score of 0.99663 (Fig. \ref{M-F1-overall}).
In comparison, $\nu$-SVR performs the best in non-anomalous with an F1 score of 0.99663. 
SVC achieved the highest accuracy overall trained cases despite not having the highest F1 score in the three test scenarios types. 

As OC-SVM is unsupervised learning, it was unable to reach the same level of accuracy compared to supervised training methods. It obtained F1 scores of between 0.87287 and 0.89189 for anomalous, non-anomalous and unstressed scenarios (Fig. \ref{M-F1-overall}). While having generally lower scores than the other models, OC-SVM did outperform the GNB model during several low-stress test cases.

For a model to be accurate in classifying the state of a system, it is essential to balance classifying both anomalous and non-anomalous scenarios. An example of an improper balance can be seen in the GNB and the OC-SVM models when looking at the icache stressor scenario, both have an accuracy of below 60\%, but their accuracy in low\_icache is above their overall accuracy (Tbl. \ref{Acc-trained}). In such cases, the boundary between anomalous and non-anomalous states is more within the non-anomalous space resulting in numerous misclassifications.

Test cases with unstable bursty behaviour regarding RTT such as aio and rawsock had lower accuracy due to creating more anomalous points within the non-anomalous stress tests. Models had difficulty classifying these RTT bursts often resulting in misclassifications during the bursts.
There are several possible reasons for these misclassification. One possible cause is that there is not enough variance in classifying RTT values in the training data.
Depending on the test, there could be a lack of data points closer to the RTT threshold resulting in some misclassifications of near threshold values. An improvement that could be tested is to include data sets that vary around the threshold between anomalous and non-anomalous. This could improve the accuracy of these intermediate results, but it could also lead to more misclassifications of non-anomalous data and more false alarms. The second is that some stressors caused a bursty nature in the RTT, which may be challenging to detect, provided the current KPI. A shorter logging interval may allow for detecting these short bursts at the expenditure of higher computational cost. 
However, there is an argument that, depending on the system,  short bursts of high RTT should not impact the overall system's stability. Furthermore, detecting anomalies that sustainably impact the systems should be the higher priority.

\begin{table}[t!]
    \fontsize{6.5pt}{7.7pt}
    \selectfont
    \centering
    \captionof{table}{Prediction accuracy results for individual trained tests}
\begin{tabular}{|c|c|c|c|c|c|c|c|c|c|c|c|c|c|c|}
 \hline
     Test & \multicolumn{2}{|c|}{GNB} & \multicolumn{2}{|c|}{$\nu$-SVR} & \multicolumn{2}{|c|}{OC-SVM} & \multicolumn{2}{|c|}{SVC} & \multicolumn{2}{|c|}{kNN} & \multicolumn{2}{|c|}{DT} & \multicolumn{2}{|c|}{RF} \\ 
      \hline
      overall & \multicolumn{2}{|c|}{83.25} & \multicolumn{2}{|c|}{99.14} & \multicolumn{2}{|c|}{79.93} & \multicolumn{2}{|c|}{99.31} & \multicolumn{2}{|c|}{99.31} & \multicolumn{2}{|c|}{99.03} & \multicolumn{2}{|c|}{99.27}  \\
      \hline
      unstressed & \multicolumn{2}{|c|}{98.49} & \multicolumn{2}{|c|}{99.33} & \multicolumn{2}{|c|}{77.48} & \multicolumn{2}{|c|}{99.16} & \multicolumn{2}{|c|}{99.33} & \multicolumn{2}{|c|}{99.16} & \multicolumn{2}{|c|}{99.33}\\ 
      \hline
      RTT Level & High & Low & High & Low &High & Low & High & Low & High & Low &High & Low & High & Low \\ 
      \hline
      aio & 98.95 & 96.84 & 98.95 & 98.74 & 98.95 & 68.2 & 98.95 & 98.74 & 98.95 & 98.74 & 98.95 & 98.11 &  98.95 & 98.74\\
      \hline
      cpu & 100.0 & 58.11 & 100.0 & 99.58 & 77.41 & 86.95 & 99.81 & 99.58 & 100.0  & 99.58 & 99.81 & 98.32 &  100.0 & 98.95\\ 
      \hline
      icache & 56.84 & 93.41 & 97.47 & 99.34 & 25.89 & 96.70 & 100.0 & 99.34 & 99.79 & 99.12 & 99.37 & 98.68 & 100.0 & 99.34 \\
      \hline
      rawsock & 100.0 & 90.30 & 100.0 & 97.23 & 99.59 & 90.89 & 100.0 & 97.03 & 100.0 & 97.03 & 100.0 & 97.03 & 100.0 & 96.83\\
      \hline
      udp & 100.0 & 19.2 & 100.0 & 99.8 & 100.0 & 58.8 & 100.0 & 99.8 & 100.0 & 99.8 & 100.0 & 99.8 & 100.0 &  99.8 \\
      \hline
    \end{tabular}
\label{Acc-trained}
\end{table}

\subsection{Untrained Results}

\begin{wrapfigure}{r}{0.5\textwidth}
    \centering
    \includegraphics[scale=0.12]{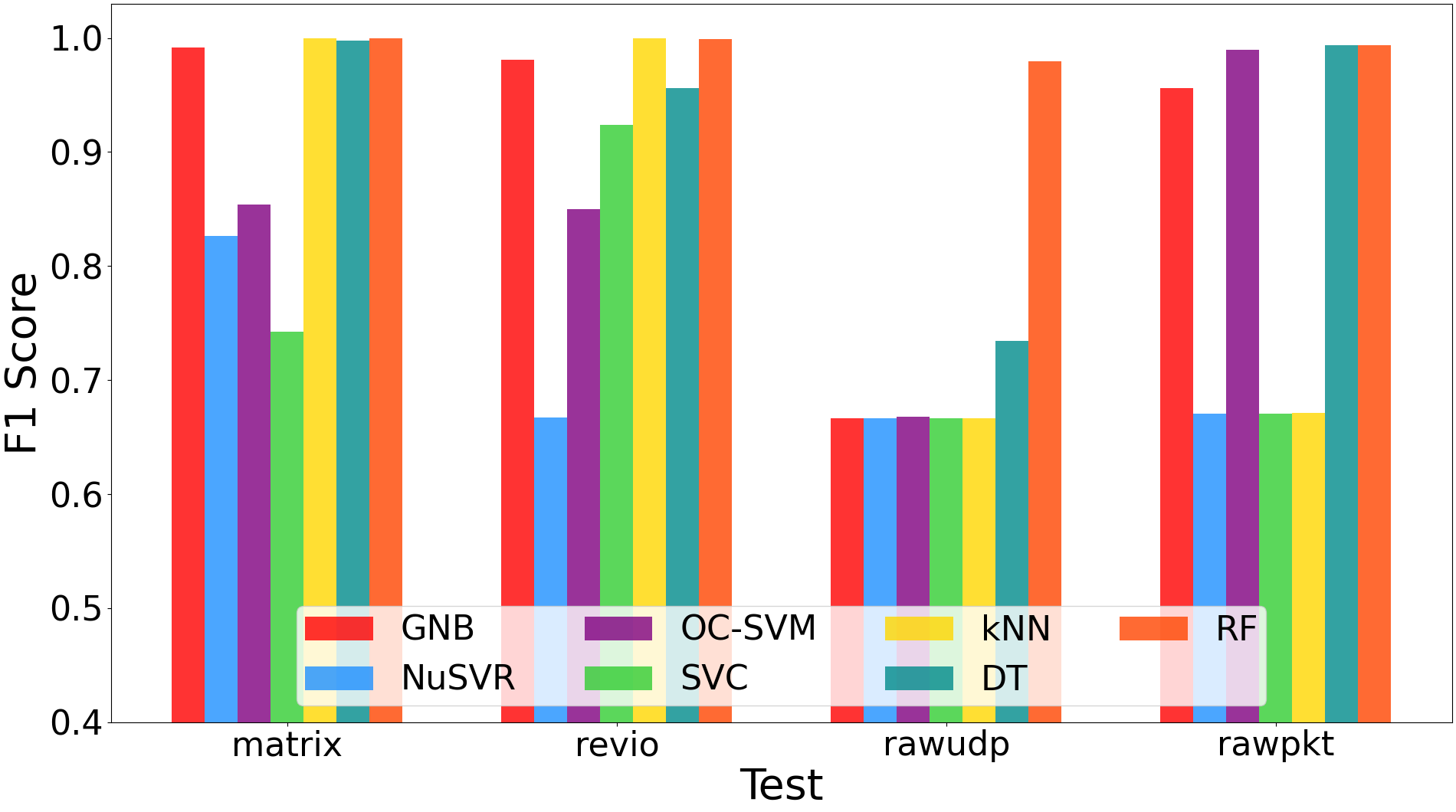}
    \caption{F1 score of models against untrained test scenarios}
    \label{M-unk-F1}
    \vspace{-10pt}
\end{wrapfigure}

The untrained results were obtained with the same test setup and method with 1000 test data points for each anomalous and non-anomalous scenario. The results showed many ML models could not classify either the anomalous or non-anomalous case for a given stressor. Both SVC and $\nu$-SVR had reduced F1-scores for all untrained stressors. The other models attained performance similar to their trained counterparts for matrix and revio test cases. The networking stressors were more challenging with DT, RF, OC-SVM, and GNB achieving good performance in rawpkt. However, RF was the only model to perform well in the rawudp scenario. Across all the untrained tests, RF achieved an F1-Score of above 0.9 (Fig. \ref{M-unk-F1}) with better accuracy regarding CPU, cache, and io interference and lower performance for the network stressors (Tbl. \ref{Acc-unk}).


The decrease in accuracy for both SVC and $\nu$-SVR could be caused by over-training. For both SVC and $\nu$-SVR, a potential cause of over-training is a high \emph{C} value creating very thin margins\cite{aurelien_handsmachine_2017}, out of the two models only SVC has a high \emph{C} value. However, when testing with lower \emph{C} values the accuracy was reduced in accuracy overall trained scenarios and the matrix and revio test cases. Revio eventually regains accuracy once reducing the \emph{C} value significantly. However, the matrix scenario then reduces in accuracy below 1\%. Another possible cause of overtraining is that higher degree polynomial kernels can cause overfitting as the shape of the support vector will mirror changes that could be caused by noise rather than fitting the overall trend of the training data \cite{mostafa_learning_2017}. The applied $\nu$-SVR does have an exponential of 7 on the polynomial kernel. However, upon testing lowering the exponential resulted in decreased accuracy for both trained and untrained CPU and io stressors with no effect on the network stressors. 
The kNN model also had poor accuracy for the tested untrained networking stressors. When performing testing the different methods, the use of Hamming distance metric greatly improves the accuracy of rawpkt and rawudp. However, it also lowers the accuracy of all other trained and untrained test scenarios with an overall accuracy of 91.316\% with higher accuracy towards non-anomalous scenarios. Switching the search algorithm to linear search had a similar impact but at an increased computation cost. Moreover, GNB and OC-SVM generally achieved comparatively higher scores despite having lower accuracy in the trained results because these models offer more flexibility at the cost of trained accuracy.

A possible cause for the lower performance of SVC, $\nu$-SVR, and kNN is that the training network stressor scenarios may have different patterns in the KPI values that cause the untrained scenarios to land on one side of the support vector. For kNN, this would instead be caused by the majority of nearest data points within the training data set being classified as anomalous. One possible way to improve these models would be to perform additional training cases covering an improved spectrum of anomalous and non-anomalous scenarios for network stressors.
RF achieved high accuracy in trained and untrained results without any overtraining regarding any of the test cases. In comparison, other models needed to give up accuracy as a trade-off for better generalization. 


\begin{table}[t!]
    \centering
    \fontsize{6.5pt}{7.7pt}
    \selectfont
    \captionof{table}{Prediction accuracy results for individual unknown tests}
\begin{tabular}{|c|c|c|c|c|c|c|c|c|c|c|c|c|c|c|}
 \hline
     Test & \multicolumn{2}{|c|}{GNB} & \multicolumn{2}{|c|}{$\nu$-SVR} & \multicolumn{2}{|c|}{OC-SVM} & \multicolumn{2}{|c|}{SVC} & \multicolumn{2}{|c|}{kNN} & \multicolumn{2}{|c|}{DT} & \multicolumn{2}{|c|}{RF} \\ 
      \hline
      RTT Level & High & Low & High & Low &High & Low & High & Low & High & Low &High & Low & High & Low \\ 
      \hline
     matrix & 99.3 & 97.4 & 100.0 & 39.0 & 55.4 & 94.8 & 99.9 & 15.7 & 100.0 & 100.0  & 99.2 & 100.0 & 99.9 & 100.0 \\
      \hline
     revio & 94.9 & 97.7 & 100.0 & 0.2 & 61.8 & 86.1 & 100.0 &  71.8 & 100.0 & 99.9 & 83.4 & 99.7 & 99.8 & 92.0\\
      \hline
      rawudp & 0.0 & 100.0 & 0.0 & 100.0 & 100.0 & 0.6 & 100.0 & 0.0 & 100.0 & 0.0 & 99.8 & 16.2 & 100.0 & 92.0 \\
      \hline
      rawpkt & 91.6 & 91.7 & 99.9 & 2.0 & 100.0 & 95.8 & 99.9 & 2.0 & 100.0 & 2.0 & 100.0 & 97.4 & 100.0 & 97.6 \\
      \hline
      \end{tabular}
\label{Acc-unk}
\end{table}

\subsection{Discussion}

The use of RTT as means of classifying training and test data is a different approach than what is presented in other ML papers that use simulated data 
\cite{jin_accurate_2016,casas_machine-learning_2016,gulenko_detecting_2018,gulenko_evaluating_2016,samir_detecting_2019,tuncer_online_2019}.
Often, the metric for anomalous is whether or not the stressor is active which for practical applications could require more verification on whether the stressor is causing the system to enter an anomalous state by exceeding the specification of the system. It could be that the system can still handle the additional load of the stressor while maintaining availability. The use of averaging and the third standard deviation of the RTT allows users to define anomalous under the perspective of meeting the system specifications and maintaining availability. From the trained results, this proved to be a viable method as all the models could achieve F1 scores above 0.8 for each scenario.

Analyzing the provided results from the trained scenarios shows that many of the ML models achieve a stable F1 score across all scenarios, excluding GNB, which has the lowest F1 score for non-anomalous despite achieving an F1 score greater than 0.95 for both anomalous and unstressed (Fig. \ref{M-F1-overall}). Among OC-SVM and GNB, there is a concern for causing false alarms for non-anomalous scenarios with higher UDP stress or detecting anomalous stress within the instruction cache as these tests had lower accuracy (Tbl. \ref{Acc-trained}). For the other support vector machine, kNN and tree-based models did not display significant levels of lower accuracy for a particular test case in the trained results. They achieved F1-scores above 0.99 for all trained scenarios (Fig. \ref{M-F1-overall}).

The untrained results presented a different perspective. They tested how well the ML models could generalize the trained results and allow the accurate detection of anomalous RTT for untrained scenarios. Out of the tested ML models, only RF achieved F1 scores above 0.97 for all test cases (Fig. \ref{M-unk-F1}). GNB, DT, and OC-SVM also have notable performances, as each showed lower performance only in the rawudp test case. $\nu$-SVR and SVC showed reduced performance across all test cases, while kNN only showed reduced accuracy during the networking test cases. From the untrained results, kNN, $\nu$-SVR, and SVC showed a greater performance decrease despite their high accuracy across the trained test cases, they proved insufficient in generalizing the approach to detecting anomalous RTT. 

The resulting applied methods show that RF will achieve the best results unless the ML is trained using anomalous training data that includes all possible anomalous cases. Using other supervised ML with highly trained accuracy could be worth creating an ensemble method with other ML methods such as DT, GNB, and OC-SVM. These models showed better performance for generalizing the trained results to create an accurate model that could detect when the systems state and provide improved assistance in maintaining availability. However, a more generalized and easy to train model could still be advantageous for self-learning applications where the model needs to completely retrain itself, which can be time-consuming and computationally intensive for large datasets.

\new{The results show that when training ML models using this methodology it is best to use a tree-based model or pair more generalized models with other ML when not able to simulate all possible loads when accuracy is of greater importance. If applying these to a telecommunication system with limited computational resources to adjust the training parameters of models such as reducing the possible depth for tree-based models or increasing the polling interval from 6s if accuracy is more important than time response.}

\section{Related Work}
\label{sec:relatedwork}

There is a growing interest in developing automated monitoring and anomaly detection for various computer systems. Many focus on different networking levels, such as core routers\cite{jin_accurate_2016} and network function virtualization\cite{sauvanaud_towards_2016}. Other papers focus on cellular networks evaluated ML techniques using data collected from active systems with anomalous data that human personnel has evaluated \cite{casas_machine-learning_2016,ciocarlie_detecting_2013}. These applications do not apply to mission-critical systems and have anomalous events lasting hours and enough data to contribute to the training process.

Regarding virtualized environments, several papers evaluate the application of anomaly detection within docker containers, and virtual machines 
\cite{gulenko_detecting_2018,sauvanaud_towards_2016}. Their application included a real-time running application that was also evaluated regarding response time. Other papers focus on modularized systems such as cluster or cloud-based services \cite{tuncer_online_2019,gulenko_evaluating_2016,samir_detecting_2019}. These applications offer metric measuring from individual virtual machines that offer modular monitoring. However, the application differs by using the data to determine whether the whole cluster or cloud is anomalous rather than a single virtualized instance. One of the papers has a similar method of training using a low and high-stress application, increasing their models' accuracy. Although it differs in that both stress levels were classified as anomalous \cite{tuncer_online_2019}.

The approach to anomaly classification differs depending on the ML application. Several papers seek to simulate hardware failures and classify anomalous as the stressor being applied\cite{gulenko_detecting_2018,casas_machine-learning_2016,samir_detecting_2019}. In this paper, the approach to anomalous classification is unique in that it is instead classified by the resulting impact on the system and whether it degrades the performance beyond the specified RTT.

\section{Conclusion}
\label{sec:conclusion}

This paper proposed a method of training ML models that can detect when given server-client communication exceeds the expected bounds by measuring only the KPI of the server side. Although different models proved to have different performance levels at detecting whether the system was in an anomalous state when provided with KPI metrics from a system loaded with a SIPp UAC with media simulations. With the provided training and test data, out of the seven models, random forests proved to be the best method to measure both the tested trained and similar untrained scenarios accurately. The evaluated methodology can provide the potential for generating anomaly detection, ensuring a more robust and reliable system for many applications and providing additional assurance for mission-critical applications.
For future work, a suggested action would be to explore the accuracy of non-binary classification in order to provide improved feedback from anomaly detection. Further testing \new{could be done to discern the most impactful KPI for training and to} validate other possible classification parameters than RTT. 

\bibliographystyle{splncs04}
{\small
\bibliography{references-trimmed}}

\section*{\uppercase{Appendix}}
\subsection{Classification Matrices and Equations}
Below are the classification matrices that were used to appropriately classify the collected data from tests as well as calculate the F1-score for determining performance.
\label{Anomaly-classification}

\begin{center}\footnotesize
\begin{tabular}{|c|c|}
    \hline
    True Positive & False Positive \\
    \hline
    \makecell{Below Anomalous RTT \\\&  Classified as \\Non-Anomalous} & \makecell{Below Anomalous RTT \\\& Classified as \\Anomalous} \\
    \makecell{False Negative} & \makecell{True Negative} \\
    \hline
    \makecell{Below Anomalous RTT \\\& Classified as \\Anomalous} &
    \makecell{Below Anomalous RTT \\\& Classified as \\Non-Anomalous} \\
    \hline
\end{tabular}
\captionof{table}{Anomaly Classification Matrix for Unstressed and Non-Anomalous Tests}
\label{non-anomalous-class}
\end{center}

\begin{center}\footnotesize
\begin{tabular}{|c|c|}
    \hline
    \makecell{True Positive} & \makecell{False Positive} \\
    \hline
    \makecell{Above Anomalous RTT \\\& Classified as \\Anomalous} &
    \makecell{Above Anomalous RTT \\\& Classified as \\Non-Anomalous} \\ 
    \hline
    \makecell{False Negative} & \makecell{True Negative} \\
    \hline
    \makecell{Below Anomalous RTT \\\& Classified as \\Anomalous} &
    \makecell{Below Anomalous RTT \\\& Classified as \\Non-Anomalous} \\
    \hline
\end{tabular}
\captionof{table}{Anomaly Classification Matrix for Anomalous Tests}
\label{anomalous-class}
\end{center}


\begin{equation} \label{eq1}
    \textit{\small Accuracy} = \textit{\small positive}/\textit{\small(positive + negative)}
\end{equation}
\begin{equation} \label{eq2}
    \textit{\small Precision} = {\textit{\small true  positive}}/({\textit{\small true  positive + false  positive}})
\end{equation}
\begin{equation} \label{eq3}
    \textit{\small Recall} = \textit{\small true  positive}/(\textit{\small true  positive + false  negative})
\end{equation}
\begin{equation} \label{eq4}
    \textit{\small F1 Score} = 2 \: \frac{\textit{\small Recall} \times \textit{\small Precision}}{\textit{\small Recall} + \textit{\small Precision}}
\end{equation}

\end{document}